\def\e{\varepsilon}
\newcommand{\lsim}{\raisebox{-.020in}{$\stackrel{<}{{\scriptstyle
\sim}}$}}
\newcommand{\gsim}{\raisebox{-.020in}{$\stackrel{>}{{\scriptstyle
\sim}}$}}
\def\sqr#1#2{{\vcenter{\vbox{\hrule height.#2pt
\hbox{\vrule width.#2pt height#1pt \kern#1pt
\vrule width.#2pt}
\hrule height.#2pt}}}}
\def\square{\mathchoice\sqr34\sqr34\sqr{2.1}3\sqr{1.5}3}

\documentstyle[aps,multicol]{revtex}

\begin{document}

\draft

\title{Entropic Barriers, Frustration and Order:
Basic Ingredients in Protein Folding}
\author{Carlos J. Camacho\\
{\normalsize \it Facultad de
F\'\i sica, P. Universidad Cat\'olica de Chile, Casilla 306, Santiago
22, Chile}\\}
\maketitle
\begin{abstract}
{
We solve a model that takes into account
entropic barriers, frustration,
and the organization of a protein-like molecule.
For a chain of size $M$, there is an effective folding
transition to an ordered structure. Without frustration,
this state is reached
in a time that scales as $M^{\lambda}$, with $\lambda\simeq 3$.
This scaling is limited by the amount of frustration
which leads to the
dynamical selectivity
of proteins:
foldable proteins are limited to
$\sim 300$ monomers; and they
are stable in {\it one} range of temperatures, independent of
size and structure.
These predictions explain generic  properties of
{\it in vivo} proteins.}
\end{abstract}
\bigskip


\begin{multicols}{2}
\columnseprule 0pt
\narrowtext

\newpage

Proteins fold to a well defined three dimensional structure, usually
referred to as the native state. However, two basic ingredients
of these biomolecules oppose this ordering process.
Namely, the large
entropy associated with the many possible conformations, and the
energetic frustration present in proteins.
{}From a physical point of view,
the interplay between order, entropy and frustration
poses some fundamental questions as to what the mechanism of the folding
process is: ``Is it possible to rationalize folding
as a relaxation process to thermodynamic equilibrium
(as some experiments suggest \cite{ANF})?'' Or ``is
some cell engineered ---e.g. chaperon
mediated \cite{ELLIS}---
mechanism needed to understand the folding process?
In this paper, we address these questions by solving
a protein-like model with all three aforementioned properties.
We find that, indeed, under some well defined conditions
kinetically foldable proteins can exist.
Moreover, the mechanism reconcile very restrictive properties
of {\it in vivo} globular proteins:
their typical size is restricted to about 300 residues (or monomers)
 and they are
stable in a {\it unique} range of temperatures!

One obstacle to folding, referred to as Levinthal's ``paradox''
\cite{LEV},
has been discussed in
several articles dealing with protein folding dynamics \cite{DILL,ZWA,CJC2}.
It relates to the fact that the
time needed to find the native state
by sampling at random the protein phase space is of the order of
the age of the universe. Nonetheless, proteins
fold in $10^{-3}$ to 1 second. Partial attempts to resolve this paradox
have been made by either suggesting some simple dynamics \cite{ZWA} or some
favorable folding conditions \cite{ES,WOL}.
Based on scaling
arguments, it has been found \cite{CJC2} that if
the ratio of hydrophobic to
hydrophilic residues in proteins is around one, then
there is a natural hierarchy in the
organization of the conformational space
of proteins. This hierarchy suggests
a {\it three stage folding kinetics} to solve the paradox. This kinetics,
sketched in Fig. 1 [9a],
has been observed in computer simulations [9b], and also seems to have
been observed in
experiments \cite{KIH}.

A second obstacle to folding is the ruggedness of the
energy landscape, or frustration.
Frustration
is particularly important when sampling
globular conformations, as in regimes II and III. Analytical
approaches to understand the role of frustration have lead to
the conclusion that random sequences of aminoacids should resemble
spin-glasses \cite{JBPW},
and therefore show glassy dynamics. Hence, it has
been argued that kinetically foldable sequences (i.e. those that
fold fast, say, in a biological time scale) must somehow
have minimum frustration [11a].

By the end of regime II,
most of the native-like structure has already been acquired.
As indicated in Fig. 1, this regime entails an entropy crisis
\cite{CJC2}
in a rugged energy landscape.
To unveil this process,
we choose
not to describe
regime III, avoiding a detailed description of the native state.
Thus, we model the folding process to a {\it generic}
native-like structure.
Given a well defined
native structure,
the microscopic model consists on (a) the definition of what a native-like
state is, (b) the characterization of the space of conformations, and (c)
the dynamics.
In what follows,
the Boltzmann constant has been set to one, and
we work in adimensional units.

The energetics involve only
short range pairwise interactions (paired monomers do not interact).
A conformation with $M=2N$ monomers has
at most $N$ non-overlapping bonds.
(a) There are $N$ native contacts defined as the set of
$N$ distinct pairs of residues
which are {\it closest} in
space in the native conformation.
All other possible
contacts are defined as non-native bonds.
Those structures
with all $N$ native contacts formed are called native-like states.
This definition does
not uniquely determine a three dimensional structure, nevertheless
this generic native-like
state is expected to resemble the overall native structure.
(b)
We consider two energy scales: $-\e_{N}<0$
for native bonds, and $\e_{NN}$ for non-native ones.
As shown in Fig. 2A, conformations are classified according to
their number of native ($i$) and non-native ($j$) bonds. Their
energy is given by
$E=-i\,\e_N
-j\,\e_{NN}$, with
$ \e_N \ge \e_{NN} \ge 0$.
The energy constant $\e_{NN}$ yields an attractive force between
non-native bonds, giving rise
to a frustrated energy landscape.
The ratio $\Delta=\e_{NN}/\e_N$ is defined as the frustration parameter
(see Fig . 2A).

To obtain the spectrum, we compute recursively
the {\it exact} crosslinking coefficient
$C(i,j)$, i.e. the number of different combinations of
$i$ native and $j$ non-native bonds among $M$ distinguishable monomers,
\begin{equation}
C(q)=\sum^q_{i,j=0} C(i,j)\,\delta_{i+j,q}={M!\over 2^qq!(M-2q)!}\,,
\end{equation}
where $q=i+j$ is the total number of bonds formed.
There is also a remanent entropy $S(q)$ associated with
the number of conformations that share
the same set of paired monomers.
For a conformation with,
say, one bond
$S(q=1)=S(l+
l_1+l_2)$, where
$l$, $l_1$ and $l_2=M-l-l_1-2q$ are the length of the loop, and
of the two free ends
of the polypeptide chain. The entropy of loops and free
ends is approximated by that of a {\it free} chain
of the same size $S_0(l)$, i.e.
$S(q)\approx
S_{0}(l)+S_{0}(l_1)+S_{0}(l_2)\equiv
S_{0}(M-2q)$, where neglecting logarithmic corrections
$S_{0}(l)=l\ln w$
and $w$ is a constant (see, e.g., Ref. \cite{CJC2}).
This assumes that
all bonded pair of monomers is equally likely,
regardless to their separation along the chain.
Hence, the whole conformational space of our model $C_T$
can be enumerated as
\begin{equation}
C_T=\sum^N_{q=0}\sum^q_{i+j=q} C(i,j)\exp[S_0(M-2q)]\,\,.
\end{equation}
Using (2), we obtain the
exact partition function, and therefore all thermodynamic
quantities.
Fig. 2A shows a scheme of the spectrum for $\Delta=1/3$, and
sketches (c) the dynamics
chosen to mimic the folding process (explained in detail in Fig. 2B).

Entropy plays a leading role on
selecting folding pathways. In particular, noting that the likelihood
of forming non-native bonds is much higher than the one of forming native
ones, there is  a natural tendency to prefer
non-native states. In this sense,
even with no energy barriers ($\Delta=0$), i.e. every single
state is connected to the ground state by an energy decreasing
pathway, non-native states form {\it entropic barriers} to folding
\cite{CJC4}!

The model has
a zero temperature transition. As shown in Fig. 3A,
the specific heat diverges (logarithmically) as $M\rightarrow\infty$.
However, proteins are finite, and
for a finite size
chain, there is an {\it effective folding transition} at $T=T_f(M)$
(defined
by the peak in the specific heat).
The inset in Fig. 3A shows that for
$T > T_f$ most bonds are non-native,
whereas for $T< T_f$ the
protein orders into a native-like state \cite{COM}.
The best fit for $T_f(M,\,\e_N,\,\Delta,\,w)$ and $M=40-1600$, $\e_N=3$,
$\e_{NN}=0-3^-$, and $w=1-5$ yields
\begin{equation}
T_f(M,\,\e_N,\,\Delta)
=1.01\,\e_N(1-\Delta)/\ln(M)+a/M,
\end{equation}
where $a$ ($\lsim\, 2$) depends on $w$ and $\Delta$, and is used as
a fitting parameter for the leading correction-to-scaling term.
For the aforementioned range of parameters, (3)
deviates from the exact values of $T_f$ (measured
with four significant figures) by less
than 1\%!
Note that the leading term in
(3) is independent of $w$. This can be understood because the
folding pathways are mostly determined by the
crosslinking coefficient $C(i,\,j)$, whereas $w$ acts as a non-specific
entropic weight.

What happens with the dynamics? The time scale to reach
equilibrium $\tau$ is measured by fitting the
exponential decay ($\exp(-t/\tau)$) of the
long-time deviation from equilibrium of any correlation function.
The time $t$ is
measured in updates of the master equation defined by the
transition probabilities in Fig. 2B. Time scales are
independent of the initial condition.
To unveil the role of the
entropic barriers we first calculate $\tau$ with no
frustration ($\Delta=0$). As shown in the inset of Fig. 3B, the equilibrium
relaxation time near $T_f$ diverges as $M\rightarrow\infty$. The divergence
of the peak in $\tau$
scales as
$\tau_c\sim M^z$,
with $z=3.8\pm0.25$ (not shown).
An striking observation is that the relaxation
time to the native-like state ($T<T_f$) scales as $\tau_0\simeq 0.45 M^\lambda$
with $\lambda=3.02\pm 0.02$,
independent of $w$
and temperature!

On adding frustration, a {\it frustration limited folding time scale}
$\tau_\Delta$ enters into the problem.
Hence, as shown in Fig. 3B,
below some temperature $T_\Delta$ the ordered state is achieved
in a time scale that diverges as $T\rightarrow 0$ as
\begin{equation}
\tau_\Delta\simeq
0.5\,M^{\lambda}\exp(2\,\e_{N}\Delta/T)/w^{4.0}\,\,\,,
\end{equation}
with $\lambda=3.14\pm 0.07$.
This expression combines both entropic barriers and
the largest minimal energy barrier
$2\,\e_{N}\Delta$ on the landscape (see Fig. 2A).
The dependence on $w$ is mostly due to the normalization factor
of the transition probabilities ---likely an artifact of the
dynamics.

For $\Delta=1/3$, the peak in $\tau$ diverges
with an exponent $z=4.3\pm 0.2$ (not shown).
More importantly, we find
a small range of temperatures around $T_\Delta=0.3$ where
---independently of size---
the native-like state is reached in a time scale $\sim\tau_0$.
Away from this temperature,
proteins will either fold too slow ($T<T_\Delta$) or
they will not fold at all ($T>T_f$). As far as we know,
this is the first time a model has predicted the dynamical
selection of a whole class of proteins in a well defined
range of temperatures!
This behavior is robust, however, the size
of the stability region and $T_\Delta$ have a {\it smooth} dependence
on $\Delta$ and $w$.
By further increasing
$\Delta$,
$\tau_\Delta$
takes over
the relaxation dynamics, even at $T>T_f$. Fig. 3B shows that already
for $\Delta =2/3$ there is no dynamical evidence
for the ordering transition at $T_f(M)$.
Substituting (3) $T=T_f$ in (4) yields
the frustration limited time scale
at the transition
\begin{equation}
\tau_\Delta^c\simeq 0.5\,M^{\zeta}/w^{4.0}\hskip0.2cm
\hbox{with}\hskip.2cm \zeta=\lambda+2\Delta /(1-\Delta).
\end{equation}
Thus, a native-like conformation is reached in a folding time
scale $\tau_f \approx \max\{\tau_\Delta^c,\, \tau_0\}$. Fast folding
sequences will fold in $\tau_f\approx\tau_0$!
Eq. 5 embodies the expected divergence of
the folding
time in the ``glassy'' limit $\Delta\rightarrow 1$. Indeed, as shown in
Fig. 4, a small change in $\Delta$ can lead
to an enormous increase in $\tau_f$.

Our results
are in excellent agreement with simulations of
random sequences of protein-like chains, where it has been shown that few
sequences can fold,
while most sequences do not ---not even to
native-like intermediates \cite{ES,ESAGMK}. Fast folding in a
limited range of temperatures has also been
observed in Monte Carlo simulations of lattice models of proteins
\cite{CJC4,JO}, and has been
suggested by Wolynes and collaborators [8,11a] in the context of
a glass transition.

It can be argued that one update of the master equation should roughly
correspond to $t_0=10^{-7}$ sec., i.e. time scale
to diffuse over some few residues
length.
This leads to the conclusion
that, even when there is no frustration, folding in a biological
time scale of seconds is restricted to globular
proteins with 300 or less
 residues (see Fig. 4).
Smaller proteins with $M\approx 40$ may fold as fast as $10^{-3}$ sec.
With frustration, proteins fold fast if
$M\,\lsim \, 300$ for $\Delta=1/3$,  and $M\,\lsim \, 20$ for $\Delta=1/2$.
These limited sizes and time scales are reasonable when compared to those
found in nature.
Hence, we conclude that the typical ratio of free energies of a
random (or non-native) bond  to that of a
native bond
must be close to $1/3$.
We note that $\Delta=1/3$ is compatible with
estimates for the relative free energies of non-bonded
interactions in proteins \cite{JER}.

The overall relaxation must also involve regime III.
The average time to escape
from one low-energy native-like state to another will depend on the energy
barrier separating them.
The structural rearrangement of a native-like state may involve
freeing a loop or, at most, a surface from its neighbors. Accordingly,
the number of broken bonds should scale as
$N^{1/3}$ for a loop, or $N^{2/3}$ for a surface. An
estimate for the escape time can then be
$t_0\exp[\e_{NN}(aN)^{1/2}/T]$ sec \cite{DT},
smaller than $\tau_0$ for numbers like $a=1/3$ and $T=0.4$.
This analysis points out to the conclusion that acquiring the native-like
structural features is the rate limiting step of the folding process.

We have solved a {\it non-sequential} model of protein folding
that focus on the rearrangement
of random conformations to close-to-native structures.
The most striking predictions are that,
regardless of the details of the
model and native structure,
folding is limited to  a well defined
range of temperatures ---$T_\Delta\, \gsim\, T\, \gsim\, T_f$ and $\e/k_BT\sim
\hbox{3-10}$---
and to globular proteins with
$M\approx 300$ or less residues.
Away from these limits, proteins do not fold.
The model
predicts a small (logarithmic) excess of heat at the ordering transition $T_f$.
We expect $T_f$ and $T_\theta$ to be rather
close for kinetically foldable proteins [9b]. Hence,
experimentally, it may
be difficult to resolve the excess of heat from these two transitions.
Entropic barriers determine
the time needed to find
a native-like state, which scales as $M^\lambda$, where $\lambda\simeq 3$.
Based on polymer dynamics insights,
a similar exponent has already been predicted for the second stage
of the folding kinetics \cite{CJC2,CJC3,DT}.
The dynamics
is governed by a multiplicity
of folding pathways with non-native-like transients.
The limitation to the aforementioned scaling time is
the amount of frustration $\Delta$, defined as the relative (attractive)
strength of non-native and native bonds.
If the amount of frustration is too large ($\Delta> 1/3$),
the time scale
to reach the native-like state scales as $M^\zeta$, where $\zeta
=\lambda + 2\,\Delta/(1-\Delta)$.
In this case, folding in a
biological time scale is slow and restricted
to very small protein sizes. We conclude
that the acquisition of native-like structure is the rate limiting
step of folding.
Folding can be rationalized based on
thermodynamic stability, at least that of native-like states.
Whether overcoming the rather large energy barriers needed to
differentiate these states requires the mediation of, say, chaperones remains
uncertain.

We thank R. Baeza, N. Bralic, and V. Tapia for stimulating
discussions. This work was supported in part by FONDECYT No. 3940016, and
DIPUC.


%

\vskip4cm

FIG. 1.
Scheme of the three stage folding kinetics.
I.- Starting from a fully unfolded
conformation R,
the initial regime
corresponds to a fast down-hill energy minimization process, where the
ruggedness of the energy surface plays almost no role.
II.-
Here, the mostly non-native
contacts ---i.e. bonds not present in the native state N---
rearrange into native ones, reaching
fairly stable and compact native-like
structures.
III.- The third
regime corresponds to the search of the native state among a
small set of close-to-native conformations separated by
rather large energy barriers.
These barriers are mostly
due to the cooperativity required by excluded volume interactions
to change
structural features buried in
the protein core.\\

\vskip8.5cm

FIG. 2.
(A) {\it Spectrum}.
Each level
$[i,\, j]$ includes all possible distinct
conformations with $i$ native and $j$ non-native number of
bonds. Some typical numbers for $M=100$ and $w=5$ are shown in parenthesis.
The slope $\Delta$ measures the amount of energetic frustration
in the model. Frustration is maximum when $\Delta=1$.
As indicated by the solid and dotted lines,
the spectrum already suggests
a multiplicity of pathways for  connecting the states.
Note, e.g., the minimum energy barrier pathway connecting
states $[0,\,N]$ and $[N,\,0]$, which
crosses over
$N-1$ small barriers of size $\e_{NN}$ and one of size $2\e_{NN}$.
(B) {\it Transition probabilities} for states $[i,\, j]$.
{}From each state $[i,\,j]$ one can
either form a new bond (non-local event) or break one (local event).
As expected in a real physical situation,
the entire temperature dependence is on
the backward transition
probabilities, and
comes from the energy penalty of breaking a bond.
The forward
probabilities, however, depend on the likelihood of forming a new bond.
Given that there
are $q$ bonds formed, the free sites can form
$M-2q\choose 2$ new bonds. $P_N$
($P_{NN}=1-P_N$) is
the probability of forming a native bond.
Detailed balanced requires an extra factor $k(q)$
proportional to the ratio of the number of conformations with $q+1$ bonds
and that with $q$ bonds. The probabilities are uniformly normalized
such that the largest transition probability of the master equation
is one half.\\

\vskip9cm

FIG. 3.
(A) {\it Thermodynamics} for $\e_N=3$, $\Delta=1/3$, and $w=5$.
Specific heat (fluctuations of the reduced
energy $\bar E=E/T$) as a function of temperature, for $M=24$,
50, 100, 200, 400, 800, and 1600. The dashed line shows the power-law
divergence of the specific heat peak
as $T_f(M)\rightarrow 0$. The inset shows the
average number of native and non-native
bonds as a function of temperature. Axes are in adimensional units, and
data is exact.
Dotted line indicates $T_f$ for $M=50$.
(B) {\it Dynamics}.
Scaled relaxation time for $\e_N=3$ and $w=5$ as a function of temperature.
Inset shows the case $\Delta=0$: For $T\,\lsim\,0.4$, the data for all values
of $M$
collapse to a constant (see $\tau_0$ in text).
Main figure shows the cases
$\Delta=1/3$ and $\Delta=2/3$.
Solid lines correspond to Eq. 4.
The horizontal axes for $\Delta=2/3$ has been rescaled by a factor 1.15.
The symbols denote $M=100$ $\times$, 80 $+$, 50 $\triangle$, 30 $\square$,
20 $\diamond$, and 10 $*$.
Error bars on $\tau$ are less
than 0.4\%.
Dotted lines
are a guide to the eye.\\

\vskip4cm

FIG. 4.
Scaling of the relaxation time to a native-like state as a function
of size and frustration.
Dotted line corresponds to the case with no frustration.
Dashed line indicates a correspondence between time
measured in updates and seconds.

\end{multicols}


\begin{references}
\bibitem{COM} The model does not recover the
coil-to-globular transition.
Even at high temperatures conformations are globular, with more
than half of its monomers paired.
One could, in principle, mimic
this transition by fitting a more appropriate entropic weight $w=w(q)$.
\end{references}
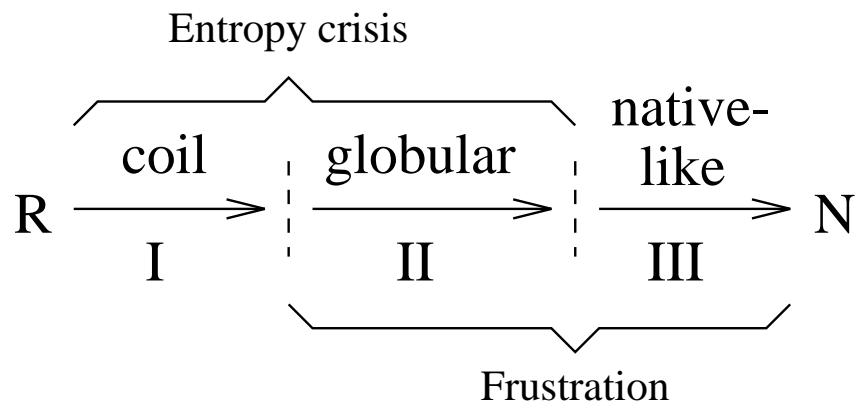
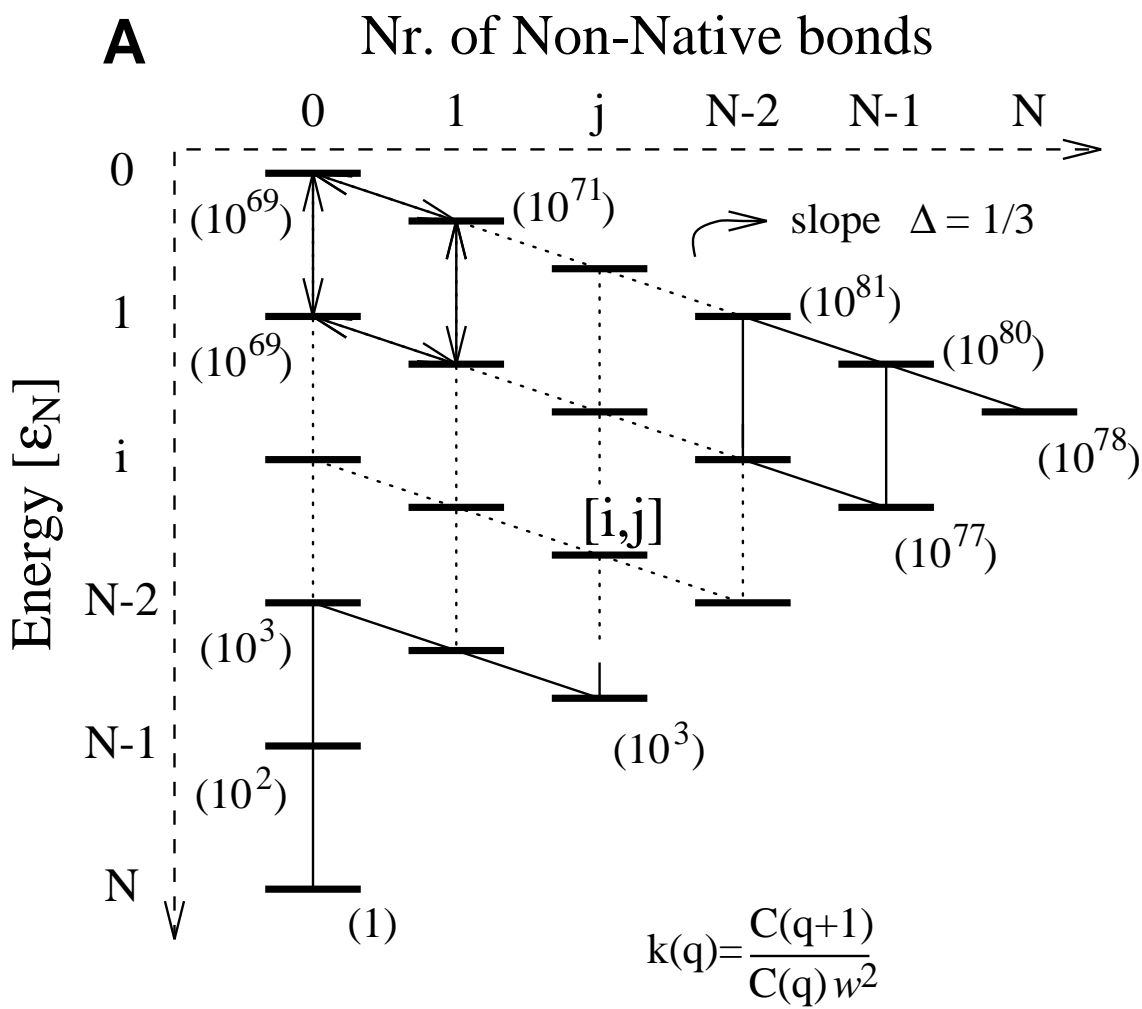
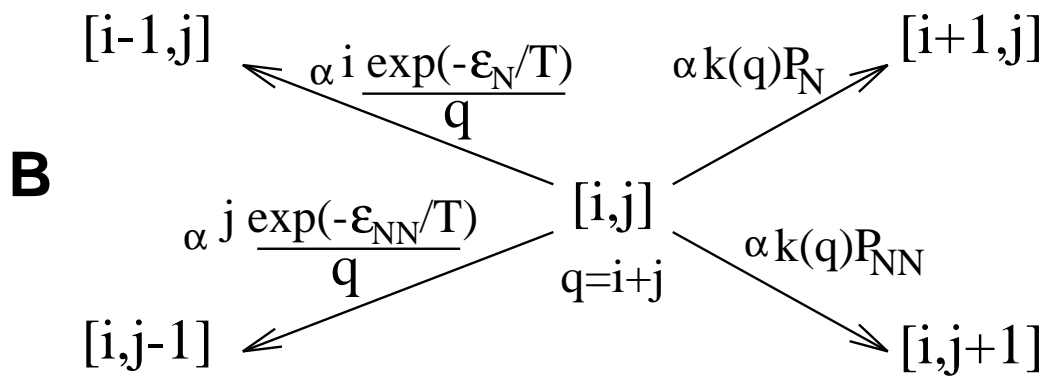
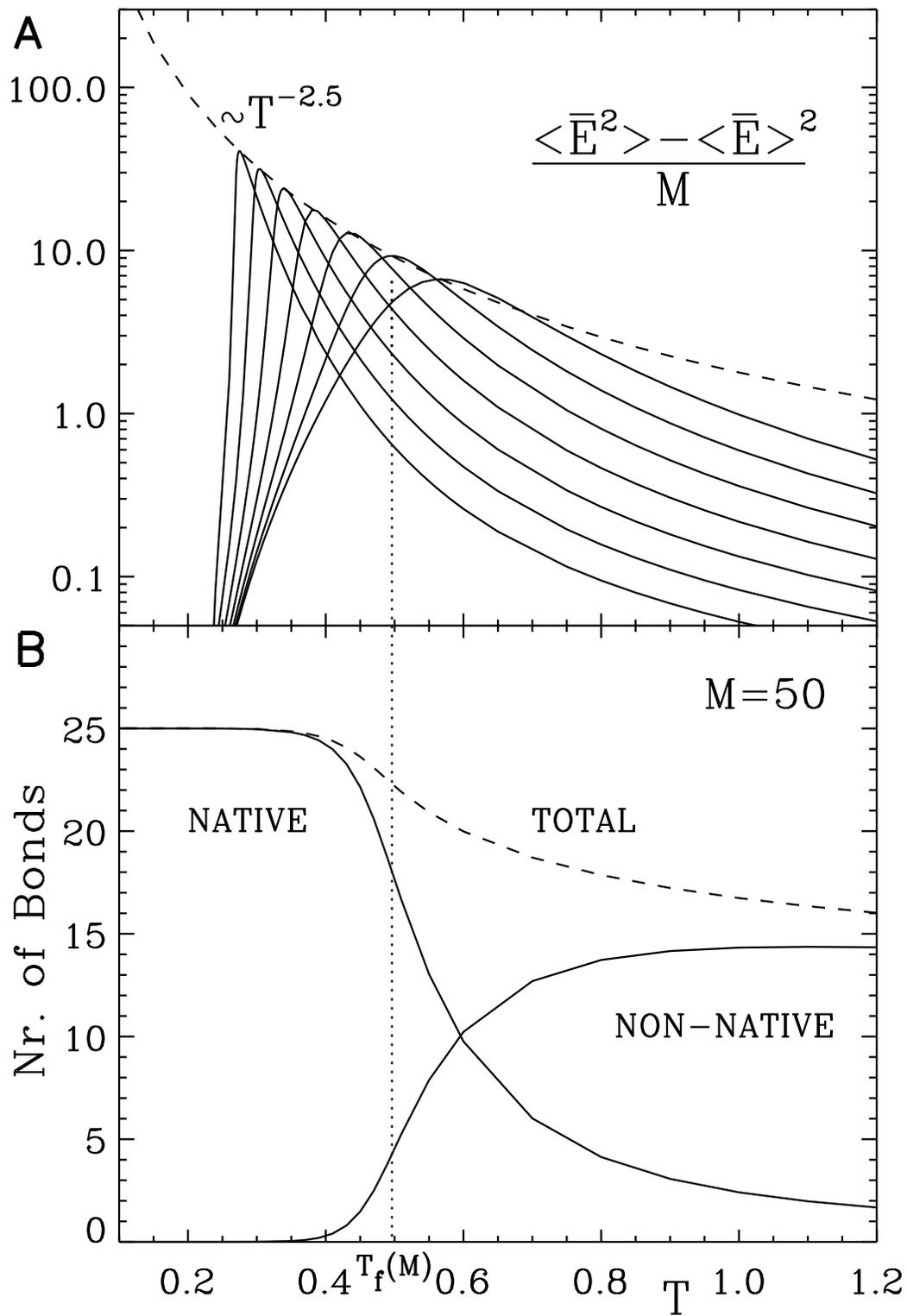
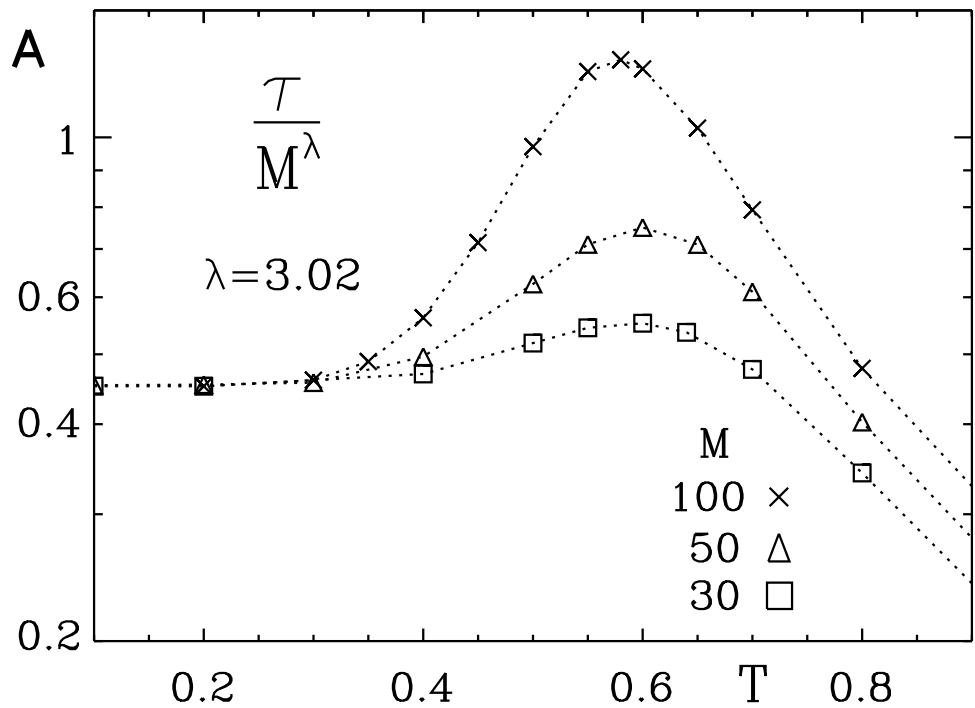
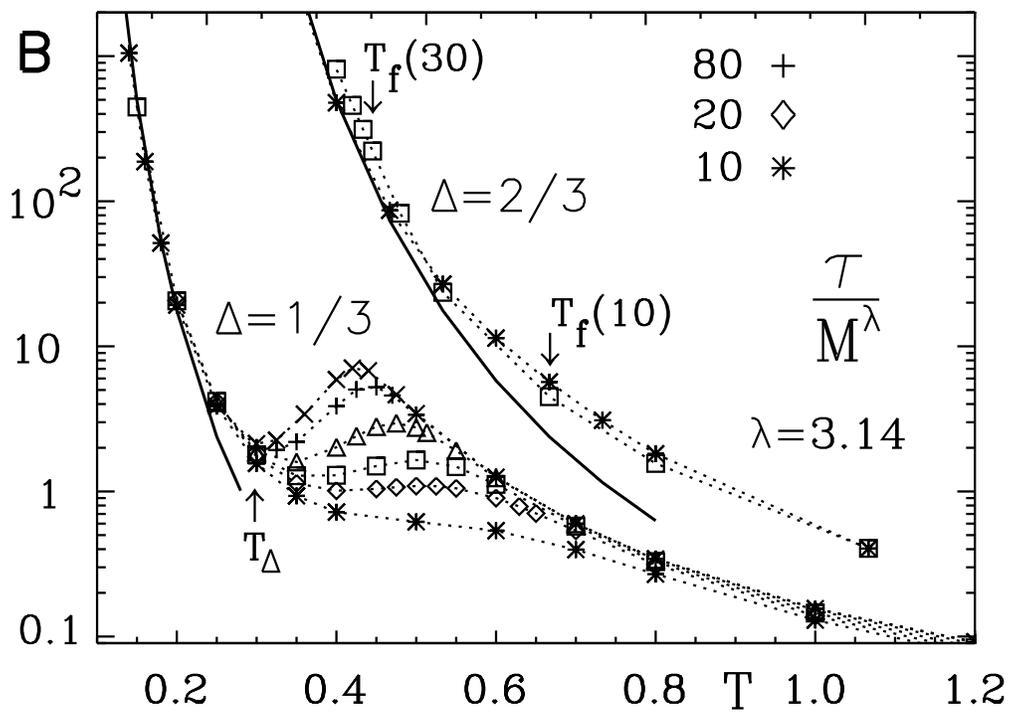
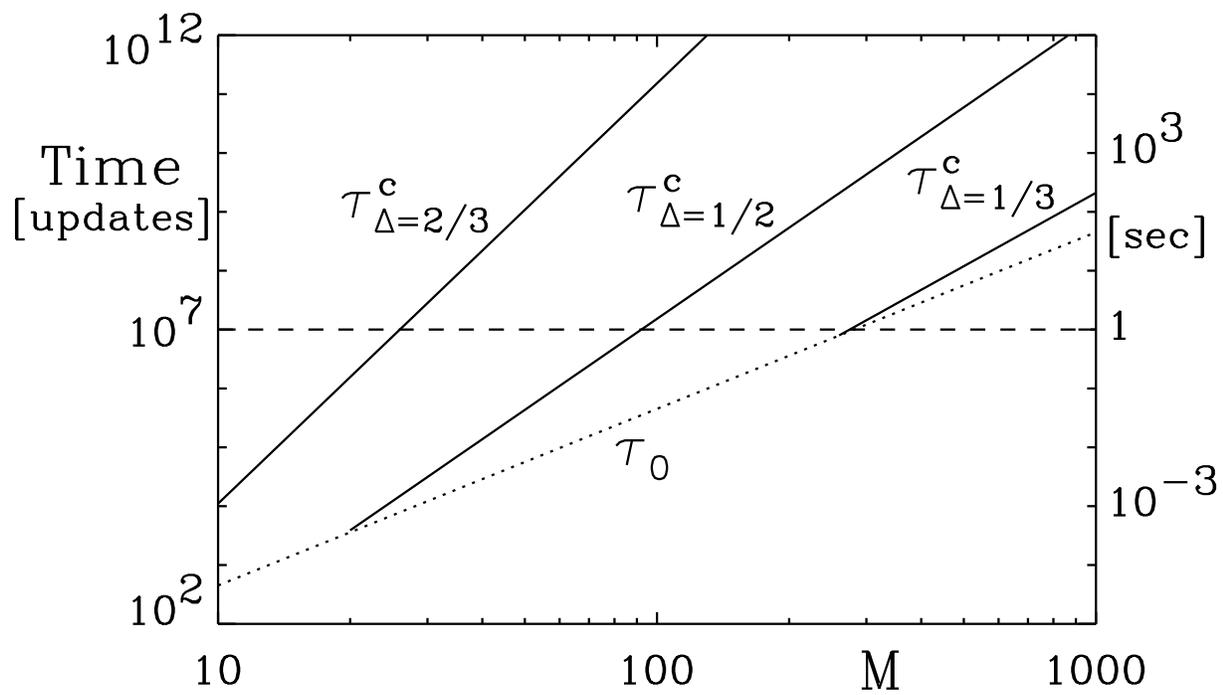
\end{document}